\documentclass[twocolumn,amsmath,showpacs,floatfix]{revtex4}
\usepackage{graphicx}

\newcommand{\degree}{\ensuremath{^{\circ}}}
\begin{document}

\title{Limits on new long-range nuclear spin-dependent forces set with a K - $^{3}$He co-magnetometer}
\author{G. Vasilakis}
\author{J. M. Brown}
\author{T. W. Kornack}
\author{M. V. Romalis}

\begin{abstract}
A magnetometer using spin-polarized K and $^3$He atoms occupying the
same volume is used to search for anomalous nuclear spin-dependent
forces generated by a separate $^3$He spin source. We measure
changes in the $^3$He spin precession frequency with a resolution of
18 pHz and constrain anomalous spin forces between neutrons to be
less than $2 \times 10^{-8}$ of their magnetic or less than $2\times
10^{-3}$ of their gravitational interactions on a length scale of 50
cm. We present new limits on neutron coupling to light pseudoscalar
and vector particles, including torsion, and constraints on recently
proposed models involving unparticles and spontaneous breaking of
Lorentz symmetry.

\end{abstract}

\affiliation{Department of Physics, Princeton University, Princeton,
New Jersey 08544}


\maketitle

Experimental limits on long-range spin-dependent forces mediated by
particles other than the photon were first considered by Ramsey
\cite{Ramsey}. Following his limit on anomalous spin forces between
protons, constraints have been set on non-electromagnetic spin
forces between electrons \cite{Ni,Bobrakov} and electrons and nuclei
\cite{Wineland}. Indirect laboratory limits on spin-dependent forces
between nuclei have been set from tests of gravitational
interactions \cite{Krauss, Adelberger} and astrophysical
considerations have been used to constrain them \cite{Engel,Haxton}.
However, no direct laboratory searches for anomalous neutron
spin-dependent forces have been performed until recently
\cite{Glenday}. Laboratory limits on anomalous forces were recently
reviewed in \cite{Adelbergerrev}. On the theoretical side, in
addition to the original motivation for spin-dependent forces
mediated by axions \cite{Moody}, a number of new ideas have been
explored, including para-photons \cite{Dobrescu}, unparticles
\cite{Georgi} and theories with spontaneous Lorentz violation
\cite{Nima}.

Here we use a co-magnetometer consisting of overlapping ensembles of
K and $^3$He atoms  to search for an anomalous interaction with a
spin source consisting of a dense nuclear spin-polarized $^3$He gas
located approximately 50 cm away. The co-magnetometer arrangement
cancels sensitivity to ordinary magnetic fields \cite{se}. After
several weeks of integration we obtain a sensitivity of 0.6~aT to an
anomalous field affecting only neutrons. For the first time, the
spin-dependent $1/r$ potential between particles is constrained
below the strength of their gravitational interactions. Our
experiment is about 500 times more sensitive and constrains more
parameters than a similar recent experiment searching for anomalous
neutron spin-dependent forces with a $^3$He-$^{129}$Xe maser that
was published after submission of this work \cite{Glenday}.

\begin{figure}[tbp]
\centering
\includegraphics[width=6.5cm]{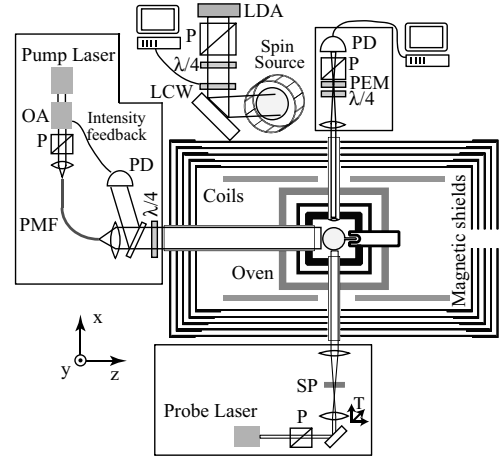}
\caption{Experimental setup. PD: photodiode, SP: stress plate to
control polarization of the probe beam, T:  translation stage to
shift the probe beam, P: polarizer, PMF: polarization maintaining
fiber, OA: Optical Amplifier, LCW: Liquid Crystal Waveplate, PEM:
Photoelastic Modulator, $\lambda$/4: quarter-waveplate, LDA: Laser
Diode Array.}\label{fig_setup}
\end{figure}

The experimental setup is shown in Fig.~\ref{fig_setup}. The
operating principle of the K-$^{3}$He co-magnetometer has been
described elsewhere \cite{se,gyro}. Briefly, the atoms are contained
in a 2.4 cm diameter spherical cell made from aluminosilicate glass
filled with 12 amagats of $^3$He, 46 Torr of N$_2$ for quenching and
a small drop of K metal. The cell is heated to 160\degree C and is
placed inside five layer $\mu$-metal shields with a shielding factor
of $10^6$. K atoms are optically pumped with a circularly polarized
pump beam generated by an amplified DFB laser. Spin-exchange
collisions between K and $^{3}$He atoms polarize $^{3}$He spins. The
current in the optical amplifier is adjusted with a slow feedback
loop to maintain a constant $^3$He polarization of about 3\%. Coils
inside the magnetic shields cancel residual magnetic fields and
create a field in the $\hat{z}$ direction parallel to the pump beam
to compensate for the effective magnetic field experienced by K
atoms due to nuclear spin magnetization of $^{3}$He. As a result,
the K magnetometer operates in a zero  field, where Zeeman resonance
broadening due to spin-exchange collisions between alkali-metal
atoms is eliminated \cite{Hap}. The polarization of K atoms in the
$\hat{x}$ direction is determined from measurements of optical
rotation of a 0.8 mW linearly polarized off-resonant probe beam
generated by a DFB laser tuned to 769.64 nm.  To achieve angular
sensitivity of $7 \times 10^{-8}$ rad/Hz$^{1/2}$ down to 0.1 Hz beam
motion due to air currents is minimized by enclosing all optics in
nearly air-tight boxes. The probe beam is carefully directed through
the center of the spherical cell to eliminate polarization rotation
caused by linear dichroism associated with reflection from tilted
surfaces.
\begin{figure}[tbp]
\centering
\includegraphics[width=8cm]{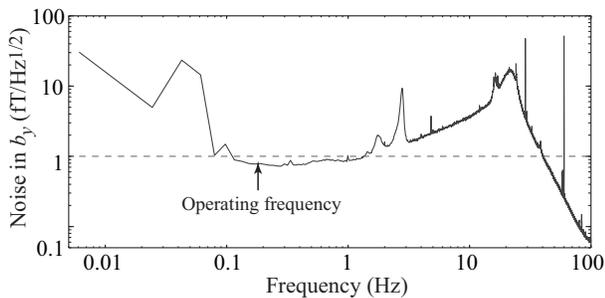}
\caption{Frequency spectrum of the co-magnetometer noise in
measurement of $b^{n}_{y}$.  The broad peak at 20 Hz is the
resonance of the coupled spin ensemble,  peaks at  2 and 3 Hz are
due to mechanical resonances of the floating optical table, and the
peak at 0.05 Hz is due to the cycle time of a cooling
system.}\label{fig_FFT}
\end{figure}

After eliminating residual magnetic fields and lightshifts using
zeroing routines described in \cite{gyro}, the $\hat{x}$
polarization of K atoms to leading order is given by
\begin{eqnarray}
P^{e}_x = \frac{P^{e}_{z}\gamma_{e}}{R_{tot}}
\left(b^{n}_{y}-b^{e}_{y}+\frac{\Omega_{y}}{\gamma_{n}}\right).
\label{signal}
\end{eqnarray}
Here $b^{n}_{y}$ and $b^{e}_{y}$ describe the phenomenological
magnetic-like fields in the $\hat{y}$ direction  that couple only to
$^{3}$He nucleus and K electrons respectively. $P^{e}_{z}$  and
$R_{tot}$ are the K electron spin polarization and relaxation rate,
$\gamma_{e}$  and $\gamma_{n}$ are the gyromagnetic ratios for
electrons and $^{3}$He nuclei respectively. Since K and $^3$He atoms
occupy the same volume, the co-magnetometer is insensitive to
ordinary magnetic fields ($b^{n}_{y}=b^{e}_{y}$) but retains
sensitivity to an anomalous field that only interacts with nuclear
spins. Previous limits on neutron-electron spin coupling
\cite{Wineland} are three orders of magnitude below our sensitivity.
$\Omega_{y}$ is the angular rotation frequency of the apparatus
relative to an inertial frame, providing an example of an
interaction that does not couple to spins in proportion to their
magnetic moments. We verified the calibration of the co-magnetometer
to 10\% accuracy by inducing small rotations of the optical table.
We also verified that the co-magnetometer is insensitive to
quasi-static magnetic fields in all directions, with the worst
suppression factor equal to $6\times 10^{-4}$ in the $\hat{z}$
direction. A typical noise spectrum of the co-magnetometer for
$b^{n}_{y}$ field is shown in Fig.~\ref{fig_FFT}. The sensitivity is
equal to 0.75 fT/Hz$^{1/2}$ at the 0.18 Hz modulation frequency of
the spin-source.

%These fields contribute to the Hamiltonian terms of $-\mu_{e}b^{n}_{y}$ and $-\mu_{^{3}He}b^{e}_{y}$, where $\mu_{e}$ and $\mu_{^{3}He}$ the electron and helion ($^{3}He$ nucleus) magnetic moment.

The anomalous field that the co-magnetometer measures is created by
optically pumped $^{3}$He nuclear spins. A cylindrical cell with 4.3
cm ID and 12.8 cm length is filled with K, $20$ Torr of N$_{2}$ and
12 atm of $^{3}$He at room temperature. The cell is heated to 190
\degree C and held in a magnetic field of 7.8 G. A broad-area laser
diode array tuned to the D1 K resonance with external grating
feedback is used for optical pumping, delivering approximately 2 W
of power to the cell. The nuclear spin direction is reversed every
2.8 sec by Adiabatic Fast Passage (AFP) using a combination of
amplitude ramp and frequency sweep of a transverse oscillating
magnetic field. With a maximum oscillating field amplitude of 0.5 G
and total sweep time of 80 msec, we achieve AFP losses of less than
$2.5 \times 10^{-6}$ per flip. A liquid crystal waveplate reverses
the direction of circular polarization of the pump beam
synchronously with the direction of nuclear polarization. Nuclear
polarization is measured using the frequency shift of the  Zeeman
resonance in the spin source correlated with $^{3}$He spin reversals
\cite{polarization}. At steady state during data acquisition, the
polarization of the spin source was 15\%, corresponding to $9 \times
10^{21}$ fully polarized $^3$He atoms. The co-magnetometer cell was
located  $48.7$ cm away from the center of the spin source cell in
the direction with altitude of $-25$\degree and azimuth of
$222$\degree.

A solenoidal coil wound on the surface of $^3$He cell along its
entire length generates a magnetic field pattern similar to that of
uniformly polarized $^3$He. We measure the magnetic field close to
the cell with a fluxgate magnetometer and adjust the current in the
coil, which is reversed synchronously with AFP flips, to reduce the
magnetic field correlated with spin reversals by a factor of 10. By
running a much larger current in the solenoid, we estimate the
leakage of the magnetic field of the spin source into the
co-magnetometer signal and limit such systematic effect to be less
than $4\times10^{-3}$ aT.

Systematic effects can also arise through parasitic cross-talk
between the electronics of the spin source and those of the
co-magnetometer. We eliminate all electrical connections between
them, with time synchronization achieved by an opto-coupled signal.
Every few days we manually change the polarity of the holding field
in the spin source and rotate a quarter-waveplate in the pump beam,
which reverses the correspondence between direction of the spins and
the state of the electronics. We also occasionally flip the
direction of the spin polarizations in the co-magnetometer, changing
the sign of its signal.

\begin{figure}[tbp]
\centering
\includegraphics[width=8cm]{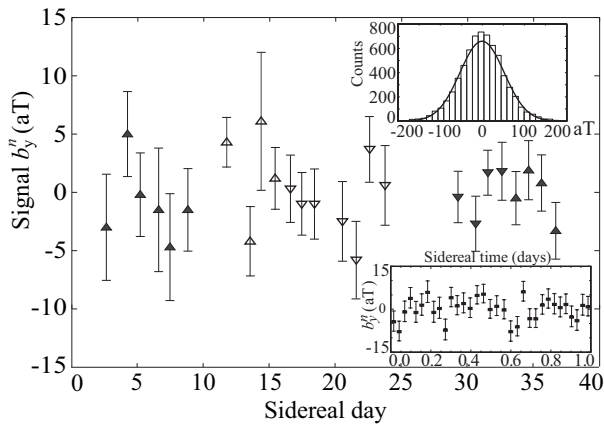}
\caption{Spin-correlated measurement of  $b^{n}_{y}$ for spin source
in the $\hat{y}$ direction. Each point represents an average over
approximately one day. Up and down triangles  indicate opposite
directions of the spin source, filled and empty triangles indicate
opposite directions of the co-magnetometer. \textit{Inset Top}:
Histogram of values for each 200 sec-long record closely follows a
Gaussian distribution. \textit{Inset Bottom}: Data plotted vs.
sidereal time of day, showing no significant
variation.}\label{fig_DotProduct}
\end{figure}

The data are collected in records of 200 sec, after which the $B_z$
magnetic field and $^3$He polarization feedback in the
co-magnetometer are adjusted, and the polarization of $^3$He in the
spin source is measured.  Approximately every 70 min automated
routines are executed to zero all magnetic fields  and the probe
beam lightshift in the co-magnetometer. The data for each record are
passed through a digital band-pass FFT filter to remove irrelevant
frequency components, the time intervals corresponding to definite
spin state are selected, and their mean and uncertainty are
calculated. An average of a 3-point moving correlation gives the
co-magnetometer signal correlated with the state of the spin source.
Fig.~\ref{fig_DotProduct} summarizes about one month of data taken
with the spin source in the $\hat{y}$ direction, oriented vertically
in the lab. The anomalous coupling $b^{n}_{y}$ is measured to be
0.05 aT $\pm$ 0.56 aT with a reduced $\chi^{2}$ of 0.87. The data
taken for different orientations of the spin source and the
co-magnetometer are consistent with each other. Measurements
performed with the spin source oriented in the $\hat{z}$ direction
give similar results $b^{n}_{y}=-0.14\pm0.84$ aT. In the $^{3}$He
nucleus, the neutron is polarized to 87\%, protons have $-2.7$\%
polarization, with the rest of the nuclear spin given by orbital
angular momentum \cite{Friar}. For simplicity we focus only on
anomalous neutron spin-dependent potential $V_a^n \sigma_n$ in the
analysis, setting $\mu_{^{3}He} b^{n}_{y} = 0.87 V_a^n$.

\begin{figure}[tbp]
\centering
\includegraphics[width=8cm]{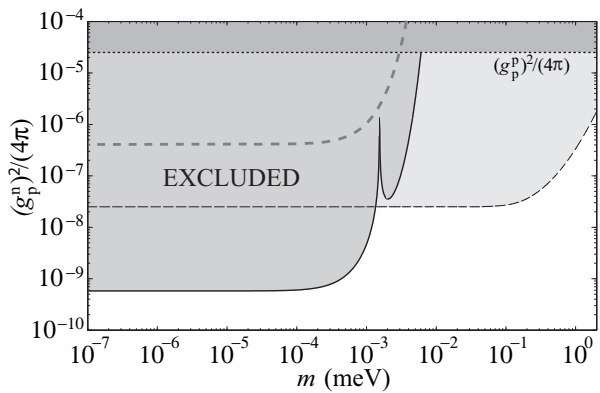}
\caption{Constraints on a pseudoscalar boson coupling to neutrons as
a function of the  boson mass. The solid line is from this work and
thin dashed line is from~\cite{Adelberger} for Yukawa coupling only.
Thick dashed line is from $^3$He-$^{129}$Xe maser~\cite{Glenday},
while the dotted line is a limit for protons set by
Ramsey~\cite{Ramsey}.} \label{figLimit}
\end{figure}

\textit{Constraints on pseudoscalar boson coupling}.  The coupling
$g_{p}$ of a pseudoscalar boson $\phi$ with mass $m$ to a fermion
$\psi$ with mass $M_n$  can be introduced using either a Yukawa or a
derivative form:
\begin{eqnarray}
\mathcal{L}^{Yuk}=-ig_{p}\overline{\psi}\gamma^{5}\psi\phi
\hspace{5.0pt}\mathrm{or}\hspace{5.0pt}
\mathcal{L}^{Der}=\frac{g_{p}}{2M_n}\overline{\psi}\gamma_\mu\gamma^{5}\psi\partial^\mu\phi.
\end{eqnarray}
Both forms lead to the same $1/r^3$ single-boson exchange potential
\cite{Moody}:
\begin{eqnarray}
V_3&=&\frac{g_{p}^2}{16\pi M_n^2}\left[\hat{\sigma}_{1}\cdot\hat{\sigma}_{2}\left(\frac{m}{r^{2}}+\frac{1}{r^{3}}\right)\right. \label{dip}\\
&-&\left.\left(\hat{\sigma}_{1}\cdot\hat{r}\right)\left(\hat{\sigma}_{2}\cdot\hat{r}\right)\left(\frac{m^{2}}{r}+\frac{3m}{r^{2}}+\frac{3}{r^{3}}\right)\right]e^{-m
r} \nonumber
\end{eqnarray}
where $r$ is the distance between the spins and $\hbar=c=1$. In
Fig.~\ref{figLimit} we show our $1 \sigma$ limit on
$(g^{n}_{p})^{2}/4\pi$ as a function of the boson mass.  For a
massless boson we obtain $(g^{n}_{p})^{2}/4\pi<5.8\times10^{-10}$, a
factor of 500 better than recent limit in \cite{Glenday}. Ramsey's
limit on proton  spin-dependent forces is
$(g^{p}_{p})^{2}/4\pi<2.3\times 10^{-5}$ \cite{Ramsey}. For a Yukawa
form of interaction, two-boson exchange leads to limits on $g_{p}$
from tests of gravitational forces,
$(g^{n}_{p})^{2}/4\pi<2.5\times10^{-8}$ \cite{Krauss,Adelberger},
but these limits do not apply to the derivative form that would be
expected for Goldstone bosons, such as the axion. There are also
astrophysical constraints on $g_{p}$ in this range from the strength
SN 1987A signal in the Kamiokande detector \cite{Engel} and
metallicity of stars \cite{Haxton}. A more reliable astrophysical
limit comes from a null search for axion emissions from the Sun at
14.4 keV M1 transition in $^{57}$Fe which constrains
$(g^{n}_{p}+0.09g^{p}_{p})^{2}/4\pi<7\times10^{-13}$ \cite{Derbin}.

\textit{Constraints on couplings to light vector bosons}.
Spin-dependent forces can also be mediated by spin-1 particles. A
para-photon that couples to fermions through dimension-six operators
is considered in \cite{Dobrescu,Dobrescu1}. It leads to a potential
similar to (\ref{dip}) but suppressed by 4 powers of a large mass
scale $M$. Our measurement constraints $M/\sqrt{ c_n}>13$~GeV,
higher than limits from electron spin-dependent forces. For a
generic dimension-four coupling of a light $Z'$ boson with mass
$m_{z'}$,
$\mathcal{L}=\overline{\psi}\gamma^{\mu}(g_{V}+\gamma_{5}g_{A}) \psi
Z'_{\mu}$,  in addition to (\ref{dip}) with $g_p^2$ replaced by
$(g_A^2+g_V^2)$, there are two more potentials \cite{Dobrescu1}:
%\begin{widetext}
\begin{eqnarray}
V_1&=&\frac{g_A^{2}}{4\pi
r}(\hat{\sigma}_{1}\cdot\hat{\sigma}_{2}) e^{-m_{z'}r} \\
V_2&=&-\frac{g_{V}g_{A}}{4\pi
M_n}\left(\hat{\sigma}_{1}\times\hat{\sigma}_{2}\right)\cdot\hat{r}\left(\frac{1}{r^2}+\frac{m_{z'}}{r}
\right) e^{-m_{z'}r}
\end{eqnarray}
Table~\ref{Bounds} summarizes the bounds from our experiment in the
limit of a massless spin-1 particle. To explicitly constrain $V_2$
we collected data with the spin source aligned in the $\hat{z}$
direction.  The constraint on $g_A^{2}/4\pi$ represents 0.2\% of the
gravitational interaction between neutrons, for the first time
constraining coupling to a massless spin-1 torsion field
\cite{Neville} below gravitational level.

\textit{Constraints on unparticle couplings to neutrons}. A new
physical entity dubbed unparticle with  unusual properties, such as
absence of a well-defined mass, has attracted a lot of attention
\cite{Georgi}. An exchange of unparticles can generate long-range
forces that vary as $1/r^{2d-1}$ where $d$ is a non-integer scaling
dimension \cite{Liao}. Spin-dependent forces are particularly
sensitive to an axial coupling of unparticles to fermions
$\mathcal{L}=C_A\overline{\psi}\gamma^{\mu}\gamma_{5} \psi
\mathcal{U}_{\mu}$. For $C_A=c_A \Lambda^{1-d}$ with $\Lambda=1$ TeV
we obtain constraints on $c_A$ shown in Table~\ref{Bounds} as a
function of $d$. These limits are similar to the ones obtained from
electron spin-dependent force \cite{Liao} and gravitational
measurements \cite{Jiang} and are much stronger than those from
astrophysics.

\begin{table}[tbp]
\begin{center}
\begin{tabular}{ccccc}
\hline \hline \multicolumn{5}{c}{Couplings to light spin-1 bosons}
\tabularnewline \hline \multicolumn{2}{c}{$V_1$: $g_{A}^{2}/(4\pi)$}
& $V_2$: $g_{V}g_{A}/(4\pi)$& \multicolumn{2}{c}{$V_3$:
$\left(g_{A}^{2}+g_{V}^{2}\right)/(4\pi)$ } \tabularnewline\hline
\multicolumn{2}{c}{$1.2\times 10^{-41}$} & $3.9 \times
10^{-26}$&\multicolumn{2}{c}{$5.8\times10^{-10}$} \tabularnewline
\hline  \hline \multicolumn{5}{c}{ Axial coupling to unparticles
with $\Lambda=1$ TeV}\tabularnewline \hline $d$: & 1 & 1.25 &1.33&
1.5 \tabularnewline \hline  $c_A$:& $1\times10^{-20}$&
$9\times10^{-16}$& $3\times10^{-14}$& $6\times10^{-11}$
\tabularnewline \hline \hline \multicolumn{5}{c}{Coupling to
Lorentz-violating Goldstone boson} \tabularnewline \hline
$M_{\pi}$(eV): & $3\times10^{-4}$& $1\times 10^{-3}$ &
$3\times10^{-3}$& $1\times10^{-2}$ \tabularnewline\hline
 $M_{\pi}/F$: & $2.1\times 10^{-20}$& $2.6\times 10^{-20}$&$2.1\times 10^{-20}$& $2.9\times 10^{-20}$ \tabularnewline \hline\hline
\end{tabular}
\end{center}
\caption{Bounds on neutron coupling to new particles.}
\label{Bounds}
\end{table}

\textit{Constraints on coupling to Goldstone bosons associated with
spontaneous breaking of Lorentz symmetry}. The dynamical effects of
a Goldstone boson $\pi$ associated with spontaneous breaking of the
Lorentz symmetry down to spatial rotations in a preferred frame have
been recently explored \cite{Nima}. Such a particle would have an
unusual quadratic dispersion relationship $\omega=k^2/M_{\pi}$ and
its leading order coupling to fermions is spin-dependent
\begin{eqnarray}
\mathcal{L}=\frac{1}{F}\overline{\psi}\gamma^\mu\gamma^{5}\psi\partial_{\mu}\pi+\frac{M^{2}_{\pi}}{F}\overline{\psi}\gamma^{0}\gamma^{5}\psi.
\end{eqnarray}
The first term gives a spin-dependent $1/r$ potential, while the
second term in a frame moving with velocity $\vec{v}$ relative to
the preferred frame  leads to an anisotropic spin interaction
$\hat{\sigma}\cdot \vec{v} $ considered in \cite{Kostelecky}.  In a
moving frame, the spin-dependent force has a complicated behavior
with a ``shock wave'' that can develop behind the spin source
\cite{Nima}. The shape of the signal  at the detector as a function
of time depends on the orientation of $\vec{v}$ relative to the
vector $\vec{r}$ from the source to the detector. For $M_{\pi} v
r>1$ the signal can average to zero over a day but has a distinctive
shape as a function of sidereal time of day. We calculated the
signal shape assuming $\vec{v}$ corresponds to the velocity of Earth
relative to the Cosmic Microwave Background radiation. The limit on
the amplitude of the signal is determined by fitting the data
plotted vs. sidereal time of day, shown in bottom inset of
Fig.~\ref{fig_DotProduct}. The bounds on $M_{\pi}/F$, shown in
Table~\ref{Bounds} for a few values of $M_{\pi}$,  reach
 below the strength of gravitational interactions. For comparison,
limits on anisotropic neutron spin interactions \cite{Walsworth}
constrain $M_{\pi}/F<4\times10^{-17}$ for $M_{\pi}=10^{-3}$ eV. We
note that our limits are in the regime where other operators in the
theory could be large and non-linear interactions in the source
could be significant. The limits can be extended to larger $M_{\pi}$
by alignment of $\vec{r}$ so $\vec{r}\cdot \vec{v}$ passes close to
$-1$, where the interaction retains its strength even for large
$M_{\pi}$.

In summary, we performed a direct search for anomalous neutron
spin-dependent forces using an alkali-metal noble-gas
co-magnetometer and a $^3$He spin source. We set limits on couplings
to several new particles, some below the strength of gravitational
interactions. We achieved an energy  resolution of $10^{-25}$ eV,
significantly higher than in other atomic physics experiments
\cite{HgEDM}, demonstrating the potential of the co-magnetometer for
future precision measurements.

 We thank J. Thaler for clarifying the
forces due to $\pi$ bosons. G. V. acknowledges assistance from V.
Papakonstantinou. This work was supported by NSF Grant No.
PHY-0653433.

\end{document}